\documentclass[letterpaper,12pt]{JHEP3}

\usepackage{graphicx}
\usepackage{amsmath}
\usepackage{amssymb}
\usepackage{xspace}
\def\hk{hyperk\"ahler\xspace}
\def\kahler{K\"ahler\xspace}
\def\hkq{/\!\!/\!\!/}

\def\bC{\mathbb{C}}
\def\bR{\mathbb{R}}
\def\bH{\mathbb{H}}
\def\bZ{\mathbb{Z}}
\def\bX{\mathbb{X}}
\def\bP{\mathbb{P}}

\def\cI{\mathcal{I}}

\def\cM{\mathcal{M}}
\def\cN{\mathcal{N}}
\def\cO{\mathcal{O}}

\def\cX{\mathcal{X}}
\def\cZ{\mathcal{Z}}

\def\frakg{\mathfrak{g}}

\def\tr{\mathop{\mathrm{tr}}}

\def\dim{\mathop{\mathrm{dim}}}

\def\vev#1{\langle#1\rangle}

\def\SU{\mathrm{SU}}
\def\U{\mathrm{U}}
\def\SO{\mathrm{SO}}
\def\SL{\mathrm{SL}}

\def\Sp{\mathrm{Sp}}
\def\SCFT{\mathrm{SCFT}}
\def\Sym{\mathrm{Sym}}
\def\adj{\mathrm{\bf adj}}
\def\omin{\cO_\text{min}}
\def\skipper{\hskip.5em\relax}
\def\Esix#1#2#3#4#5#6{%
{\text{\small$\begin{array}{c@{\skipper}c@{\skipper}c@{\skipper}c@{\skipper}c@{\skipper}c}
&&#2&& \\
#1 &  #3 &  #4 & #5 & #6
\end{array}$}}}
\def\proj#1{\bigm|_{#1}}

\let\tilde\widetilde
\let\bar\overline
\preprint{
\hbox{}\hfill arXiv:0810.4541}

\title{Argyres-Seiberg duality and the Higgs branch}

\author{

Davide Gaiotto$^{1}$, Andrew Neitzke$^2$ and Yuji Tachikawa$^1$ \\

\bigskip
$^1$ School of Natural Sciences, Institute for Advanced Study,\\
 Princeton,  New Jersey 08540, USA \\
 $^2$ Jefferson Physical Laboratory, Harvard University, \\
Cambridge, MA 02138, USA
}

\abstract{
We demonstrate the agreement
between the Higgs branches of two $\cN=2$ theories proposed by Argyres and Seiberg to be $S$-dual,
namely the $\SU(3)$ gauge theory with six quarks, and
the $\SU(2)$ gauge theory
with one pair of quarks coupled to the superconformal theory with $E_6$ flavor symmetry.
In mathematical terms, we demonstrate the equivalence between a \hk quotient of a linear space and another \hk quotient involving the minimal nilpotent orbit of $E_6$,
modulo the identification of the twistor lines.
}


\begin{document}


\setcounter{tocdepth}{1}

\section{Introduction}
\label{introduction}
\subsection{Argyres-Seiberg duality}
In a remarkable paper \cite{Argyres:2007cn}, a new type of strong-weak duality of four-dimensional $\cN=2$ theories was introduced.
Consider an $\cN=2$ supersymmetric $\SU(3)$ gauge theory
with six quarks in the fundamental representation.
This theory has vanishing one-loop beta function, and the gauge coupling constant
\begin{equation}
\tau=\frac{\theta}{\pi}+\frac{8\pi i}{g^2}
\end{equation}
is exactly marginal.
Argyres and Seiberg carried out a detailed study of
the behavior of the Seiberg-Witten curve
close to the point $\tau\to 1$ where the theory is infinitely strongly-coupled,
and were  led
to conjecture a dual description involving an $\SU(2)$ group with gauge coupling
\begin{equation}
\tau'=\frac{1}{1-\tau}.
\end{equation}

To understand the matter content of the dual theory, one first needs to recall
the interacting superconformal field theory (SCFT) with flavor symmetry $E_6$
first described by \cite{Minahan:1996fg}. This theory has one-dimensional Coulomb branch
parametrized by $u$ whose scaling dimension is 3,
and is realized as the low-energy limit of the worldvolume theory on a D3-brane
probing the transverse geometry of an F-theory 7-brane with $E_6$ gauge group.
The gauge group on the 7-brane then
 manifests as a flavor symmetry from the point of view of the D3-brane.
We denote this theory by $\SCFT[E_6]$ following \cite{Argyres:2007cn}.

Now, the theory Argyres and Seiberg proposed as the dual of the $\SU(3)$
gauge theory with six quarks consists of the $\SU(2)$ gauge bosons,
coupled to one hypermultiplet in the doublet representation,
and also to a subgroup $\SU(2)\subset E_6$ of $\SCFT[E_6]$.
The $\SU(2)$ subgroup is chosen so that
the raising operator of $\SU(2)$ maps to the raising operator for the highest root of $E_6$.
In the following, we refer to two sides of the duality as the $\SU(3)$ side and the exceptional side, respectively.

Argyres and Seiberg provided a few compelling pieces of evidence for this duality.
First, the flavor symmetry agrees. On the $\SU(3)$ side, there is a $\U(6)=\U(1)\times \SU(6)$
symmetry which rotates the six quarks. On the exceptional side, there is an $\SO(2)$ symmetry
which rotates a pair of quarks in the doublet representation, which can be identified with the $\U(1)$ part of $\U(6)$.
Then, the flavor symmetry of the SCFT with $E_6$ is broken down to the maximal subgroup commuting
with $\SU(2)\subset E_6$, which is $\SU(6)$.
Second, the scaling dimensions of Coulomb-branch operators agree.
Indeed, on the $\SU(3)$ side one has $\tr \phi^2$ and $\tr\phi^3$
where $\phi$ is the adjoint chiral multiplet of $\SU(3)$. The dimensions are thus
2 and 3. On the exceptional side, one has $\tr\varphi^2$ (where
$\varphi$ is the adjoint chiral multiplet of $\SU(2)$), which has dimension $2$, and the Coulomb-branch
operator $u$ of $\SCFT[E_6]$, which has dimension $3$.

Third, Argyres and Seiberg studied in detail the deformation of the Seiberg-Witten curve under the
$\SU(6)$ mass deformation, and found remarkable agreement.
Fourth, they computed the current algebra central charge of the $\SU(6)$ flavor symmetry
on the $\SU(3)$ side, which agreed with the central charge of the $E_6$ symmetry
on the exceptional side, inferred from the fact that the beta function of the
$\SU(2)$ gauge group coupling is zero.
This is as it should be, because $\SU(6)$ arises as a subgroup of $E_6$ on the
exceptional side.
This provided a prediction of the current central charge of $\SCFT[E_6]$
for the first time, which was later reproduced holographically by \cite{Aharony:2007dj}.
There are generalizations to similar duality pairs
involving  SCFTs with  flavor symmetries other than $E_6$ \cite{Argyres:2007cn,Argyres:2007tq}.

Our aim in this note is to present further convincing evidence for this duality,
by showing that the Higgs branches of the two sides of the duality are equivalent as \hk cones. Mathematically speaking, we will show the agreement of
their twistor spaces as complex varieties with real structure,
but we have not been able to prove that they share the same family
of twistor lines. Instead we give numerical evidence that their K\"ahler potentials agree  in Appendix~\ref{kahler}.

\subsection{Higgs branch}\label{rough}
On the $\SU(3)$ side, let us denote the squark fields by \begin{equation}
Q^i_a,\qquad \tilde Q^a_i
\end{equation} where $i=1,\ldots,6$ are the flavor indices and $a=1,2,3$ the color indices.
The Higgs branch is the locus where the F-term
and the D-term both vanish,
divided by the action of the gauge group $\SU(3)$.
As is well known, this space can also be obtained by setting $F=0$
without setting $D=0$, and dividing by the complexified gauge group
$\SL(3,\bC)$.

Thus the Higgs branch is parametrized by gauge invariant composite operators  \begin{equation}
M^i{}_j= Q^i_a\tilde Q^a_j,\qquad
B^{ijk}=\epsilon^{abc} Q^i_a Q^j_b Q^k_c,\qquad
\tilde B_{ijk}=\epsilon_{abc} \tilde Q^a_i \tilde Q^b_j \tilde Q^c_k
\end{equation} which satisfy various constraints, e.g.~\begin{equation}
B^{[ijk}M^{l]}{}_m=0
\end{equation}  to which we will come back later.
The fields $Q^i_a$, $\tilde Q^a_i$ have 36 complex components,
while the F-term condition imposes 8 complex constraints.
The quotient by $\SL(3,\bC)$ reduces the complex dimension further by 8,
so the Higgs branch has complex dimension \begin{equation}
2\times 3\times 6 - 8 - 8 = 20.
\end{equation}

Our problem is to understand how this structure of the Higgs branch is realized on the exceptional side.
Firstly, we have one hypermultiplet in the doublet representation,
which we denote as $v_\alpha$, $\tilde v_\alpha$ in $\cN=1$ superfield notation.
Here $\alpha=1,2$ is the doublet index.

We also have the Higgs branch of $\SCFT[E_6]$, the structure of which is known through
the F-theoretic construction of the SCFT. Recall that this theory is the worldvolume theory
on one D3-brane probing a F-theory 7-brane of type $E_6$.
Say the D3-brane extends along the directions $0123$, and the 7-brane
along the directions $01234567$.
The one-dimensional Coulomb branch of this theory is identified
with the transverse directions $89$ to the 7-brane.
The theory becomes superconformal when the D3-brane hits the 7-brane,
at which point the Higgs branch emanates. This is identified as the process
where a D3-brane is absorbed into the worldvolume of the 7-brane as an $E_6$
instanton along the directions $4567$.
The real dimension of the $N$-instanton moduli space of $E_6$ is $4 h_{E_6} N$
with the dual Coxeter number $h_{E_6}=12$. The center-of-mass motion
of the instanton corresponds to a decoupled free hypermultiplet,
and thus the genuine moduli space is the so-called `centered'
one-instanton moduli space without
the center-of-mass motion, which has complex dimension $22$.

The $\SU(2)$ gauge group couples to the quark fields $v_\alpha$, $\tilde v_\alpha$,
and this instanton moduli space. Imposing the F-term condition and dividing by the
complexified gauge group, we find the complexified
dimension of the Higgs branch as \begin{equation}
2\times 2 + 22 - 3 - 3 = 20,
\end{equation} which correctly reproduces the dimension of the Higgs branch
on the $\SU(3)$ side.

We would like to perform more detailed checks, and for that purpose
one needs to have a concrete description of the instanton moduli.
It is well known that the ADHM description is available for classical gauge groups,
but how can we proceed for exceptional groups?
Luckily,  there is another description of the 1-instanton moduli spaces,
applicable to any group $G$, which identifies the centered 1-instanton moduli space with the
minimal nilpotent orbit of $G$ \cite{Kronheimer}.

Let us now define the minimal nilpotent orbit.
$G_\bC$ acts on the complexified Lie algebra $\frakg_\bC$, which has the Cartan generators $H^i$
and the raising/lowering operators $E_{\pm\rho}$ for roots $\rho$.
$G_\bC$ also acts on the dual vector space $\frakg_\bC^*$ of $\frakg_\bC$ via the coadjoint action,\footnote{One can of course identify
$\frakg_\bC^*$ and $\frakg_\bC$ using the Killing form, but it is more mathematically natural to use the coadjoint representation here.}\
and the minimal nilpotent orbit $\omin(G)$ of $G$ is
the orbit of $(E_\theta)^*$, where $\theta$ denotes the highest root: \begin{equation}
    \omin(G)=G_\bC \cdot (E_\theta)^*.
\end{equation}
The minimal nilpotent orbit is known to have polynomial defining equations.
Moreover, they can be chosen to be quadratic, transforming covariantly under $G_\bC$.
These relations are known under the name of the Joseph ideal \cite{Joseph}.
The simplest example is the case $G = \SU(2)$.  In this case $\frakg_\bC$ is  three-dimensional;
denote its three coordinates by $a$, $b$ and $c$, which
transform as a triplet of $\SU(2)$.  The minimal nilpotent orbit is then given by
\begin{equation}
a^2+b^2+c^2=0
\end{equation} which describes the space $\bC^2/\bZ_2$,
and as is well-known, the centered one-instanton moduli space of $\SU(2)$ is exactly this orbifold.

Let us come back to the case of $E_6$.
We fix an $\SU(2)$ subalgebra generated by $E_{\pm\theta}$.
The maximal commuting subalgebra is then $\SU(6)$.
The $E_6$ algebra can be decomposed under
the subgroup $\SU(2)\times\SU(6)$ into
 \begin{equation}
X^i{}_j,\qquad  Y^{[ijk]}_\alpha, \qquad Z_{\alpha\beta}
\end{equation} where $i,j,k=1,\ldots,6$ are the $\SU(6)$ indices,
$\alpha,\beta=1,2$ those for $\SU(2)$.  Here $X^i{}_j$ and $Z_{\alpha\beta}$
are adjoints of $\SU(6)$ and  $\SU(2)$ respectively, and $Y^{ijk}_\alpha$ transforms
as the three-index anti-symmetric tensor of $\SU(6)$ times the doublet of $\SU(2)$.
The minimal nilpotent orbit is then given by the simultaneous zero locus of
quadratic equations in $X$, $Y$ and $Z$ which we describe in detail later.

For now let us see what are the gauge-invariant coordinates of the Higgs branch of the exceptional side.
The $\SU(2)$ gauge group is identified to the $\SU(2) \subset E_6$
just chosen above, i.e. the $\SU(2)$ gauge bosons couple to
the current of this $\SU(2)$ subgroup of the $E_6$ symmetry.
We also have the quarks $v_\alpha$ and $\tilde v_\alpha$ in addition
to the fields $X$, $Y$ and $Z$, and we need to make $\SU(2)$-invariant combinations of them.
Moreover, we need to impose the F-term equation, which is \begin{equation}
Z_{\alpha\beta}+v_{(\alpha}\tilde v_{\beta)}=0
\end{equation} as we argue later.
Thus, any appearance of $Z$ inside a composite operator can be eliminated in favor of $v$ and $\tilde v$.
Therefore we have the following natural gauge-invariant composites,
from which   all gauge-invariant operators  can be generated as will be shown in Sec.~\ref{5.4}: \begin{equation}
(v\tilde v),\qquad X^i{}_j,\qquad (Y^{ijk} v),\qquad (Y_{ijk} \tilde v).
\end{equation}
Here we defined \begin{equation}
(uw)\equiv u_\alpha w_\beta\epsilon^{\alpha\beta}
\end{equation}for two doublets $u_\alpha$ and $w_\alpha$,
and $Y_{ijk,\alpha}$ is defined by lowering the indices of $Y^{ijk}_\alpha$
by the epsilon tensor, see Appendix~\ref{conventions}.

This suggests the following identifications between the operators on the two sides of the duality:
\begin{align}
\tr M &\leftrightarrow (v\tilde v), & \hat M^i{}_j &\leftrightarrow X^i{}_j,\\
B^{ijk}&\leftrightarrow (Y^{ijk}v),& \tilde B_{ijk} &\leftrightarrow (Y_{ijk}\tilde v)
\end{align} where
$\hat M^i{}_j$ is the traceless part of $M^i{}_j$.
The identifications preserve the dimensions of the operators if we assign
dimensions 2 to the fields $X$, $Y$ and $Z$.
The $\SU(6)$ transformation nicely agrees. The $\U(1)$ part of the flavor symmetry
can be matched if one assigns charge $\pm1$ to $Q$, $\tilde Q$,
and charge $\pm3$ to $v$, $\tilde v$.

This factor of 3 was predicted in the original paper \cite{Argyres:2007cn}
from a totally different point of view, by demanding that the
the two-point function of two $\U(1)$ currents should agree under the duality.
Let us quickly recall how it was derived there. The form of the
two-point function of the $\U(1)$ current $j_\mu$ is strongly constrained by the conservation
and the conformal symmetry, and we have \begin{equation}
\vev{j_\mu(x)  j_\nu(0)} \propto k \frac{x^2 g_{\mu\nu}-2x_\mu x_\nu}{x^8}  +\cdots.
\end{equation} $k$ is called the central charge, and $\cdots$ stands for less singular terms.
Let us normalize $k$ such that one hypermultiplet of charge $q$ contributes $q^2$ to $k$.
Assign $Q$, $\tilde Q$ the charge $\pm 1$, and let the charge of $v$, $\tilde v$  be $\pm q$.
Then $k$ calculated from the $\SU(3)$ side is $6\times 3=18$, while
$k$ determined from the exceptional side is $2 q^2$. Equating these, Argyres and Seiberg concluded
that the charge of $v$, $\tilde v$ should be $q=\pm 3$.

The agreement is already impressive at this stage, but we would like to see
how the constraints are mapped. We would also like to study how the \hk structures agree,
because so far we considered the Higgs branch only as a complex manifold.
For that purpose we need to recall more about the \hk cone.

The structure of  the rest of the paper is as follows:
We discuss in Sec.~\ref{formalities}
what data are mathematically necessary to show the equivalence of the Higgs branches.
Sec.~\ref{orbit} is devoted to the description of the minimal nilpotent orbit, i.e.~the 1-instanton moduli space, as a \hk space.
Sec.~\ref{SU3} and Sec.~\ref{E6} will be spent in calculating
the necessary data on the $\SU(3)$ side
and the exceptional side, respectively. Then they are compared in Sec.~\ref{comparison}
which shows remarkable agreement.
We conclude in Sec.~\ref{conclusion}.
We have four Appendices: Appendix~\ref{conventions} collects our conventions,
Appendix~\ref{twistors} gathers the machinery of twistor spaces required to show
the equivalence of \hk cones, and
Appendix~\ref{kahler} compares the \kahler potentials of the duality pair.
Appendix~\ref{math-summary} is a summary for mathematicians.

\section{Rudiments of \hk cones}\label{formalities}
Here we collect the basics of the \hk cones in a physics language.
Mathematically precise formulation can be found in \cite{Swann,Br}.
The Higgs branch $\cM$ of an $\cN=2$ gauge theory is a \hk manifold, i.e.~one
has three complex structures $J^{1,2,3}$ satisfying $J^1 J^2 =J^3$, compatible with the metric $g$,
and the associated two-forms $\omega_{1,2,3}$ are all closed.
We choose a particular $\cN=1$ supersymmetry subgroup  of the $\cN=2$ supersymmetry group,
which distinguishes one of the complex structures, say $J\equiv J^3$.
$\cM$ is then thought of as a \kahler manifold with the \kahler form $\omega=\omega_3$.
$\Omega=\omega_1 + i \omega_2$ is a closed $(2,0)$-form on $\cM$
which then defines a holomorphic symplectic structure on $\cM$.
Physically this means that the $\cN=1$ chiral ring, i.e.~the
ring of holomorphic functions on $\cM$, has a natural holomorphic Poisson bracket
\begin{equation}
[f_1,f_2]=(\Omega^{-1})^{ij} \partial_i f_1 \partial_j f_2
\end{equation}  for two holomorphic functions $f_{1,2}$ on $\cM$.

Second, we are dealing with the Higgs branch of an $\cN=2$ superconformal theory,
which has the dilation and the $\SU(2)_R$ symmetry built in the symmetry algebra.
The dilation makes $\cM$ into a cone with the  metric \begin{equation}
ds^2_{\cM} = dr^2 + r^2 ds^2_\text{base},
\end{equation}  and $\SU(2)_R$ symmetry acts on the base of the cone as an isometry,
rotating the three complex structures as a triplet. These two conditions make
$\cM$ into a \hk cone. $K=r^2$ is a \kahler potential with respect to any of the complex structures
$J^{1,2,3}$,  and is called the \hk potential in the mathematical literature.
The dilatation assigns the scaling dimensions, or equivalently the weights, to the chiral operators on $\cM$.

Let us consider an element of $\SU(2)_R$ which acts on the three complex structures
as $(J^{1},J^2,J^3)\to (J^1,-J^2,-J^3)$.
This element defines an anti-holomorphic involution $\sigma:\cM\to \cM$
because it reverses $J\equiv J^3$.
This induces an operation $\sigma^*$ on holomorphic functions
on $\cM$ via $( \sigma^*(f) ) (x) \equiv  f(\sigma(x))$.
$\sigma^*$ maps holomorphic functions to anti-holomorphic functions,
but  is a linear operation, not a conjugate-linear operation.
We call this operation the conjugation.

As will be detailed in Appendix~\ref{twistors}, the space $\cM$  as a complex manifold,
with the Poisson brackets, the scaling weights and the conjugation, almost suffices to reconstruct
the \hk metric on $\cM$.
Therefore, our main task in checking the agreement of the Higgs branches
of the duality pair is to identify them as complex manifolds,
and to show that the extra data defined on them also coincide.
In order to complete the proof we need to show that the families
of the twistor lines coincide, which we have not been able to do.
Instead we will give numerical support by calculating the K\"ahler potential
directly on both sides in Appendix~\ref{kahler}.

The Higgs branches $\cM$ that we treat here are gauge theory moduli spaces.
They can be described by the \hk quotient construction \cite{Hitchin:1986ea},
which we now review.
Let us start with an $\cN=2$ gauge theory with the gauge group $G$,
whose hypermultiplets take value in the \hk manifold $\cX$.
The action of $G$ on $\cX$ preserves three \kahler structures, and thus there are
three moment maps $\mu_s^a$ ($s=1,2,3$; $a=1,\ldots,\dim G$)
which satisfy \begin{equation}
d\mu_s^a = \iota_{\xi^a}\omega_s
\end{equation} where $\xi^a$ is the Killing vector associated to the $a$-th generator of $G$.
The Higgs branch of the gauge theory, in the absence of any non-zero Fayet-Iliopoulos parameter,
is then given by \begin{equation}
\cM \equiv \cX\hkq G \equiv \{ x\in \cX\bigm| \mu^a_s(x)=0 \} / G.
\end{equation}
With one complex structure $J=J^3$ chosen, it is convenient to call \begin{equation}
D^a= \mu^a_3, \qquad F^a=\mu^a_1+i\mu^a_2.
\end{equation} Then, as a complex manifold, \begin{equation}
\cM = \{ x\in \cX \bigm| F^a=0 \} /G_\bC.
\end{equation}
It is instructive to note that $F^a$ is exactly the Hamiltonian which generates the $G$  action
on the chiral ring of $\cX$, under the Poisson bracket associated to $\Omega=\omega_1+ i\omega_2$.

The conjugation $\sigma^*$ and the Poisson bracket $[\cdot,\cdot]$ on the quotient $\cM$ are given by the restriction of the corresponding operations on $\cX$.  It is instructive to see why the Poisson bracket of the quotient
is well-defined: two $G$-invariant holomorphic functions $f_{1,2}$ on $\cX$ lead to the same function
on $\cM$ if and only if $f_1= f_2+u_a F^a$ with holomorphic functions $u_a$.
Then we have, for a $G$-invariant holomorphic function $h$,
\begin{equation}
[f_1,h]- [f_2,h]=[u_a F^a,h]=[u_a,h]F^a + u_a [F^a,h]
\end{equation} on $\cX$.
The first term in the right hand side is zero on $\cM$ because we set $F^a=0$,
while the second term is zero because $h$ is $G$-invariant.
Therefore  $[f_1,h]$ and $[f_2,h]$ determine the same holomorphic function on $\cM$.

The \kahler potential of $\cM$ is similarly the restriction of that of $\cX$ to the zero locus of the moment maps in our situation, as discussed in Sec.~2B of \cite{Hitchin:1986ea}.
To illustrate the procedure,
let us consider an $\cN=1$ supersymmetric $\U(1)$ gauge theory
coupled to chiral fields $\Phi_i$ of charge $q_i$ whose Lagrangian is \begin{equation}
L=\int d^4\theta\, (\sum_i\Phi_i^* e^{2q_iV}\Phi_i +\xi  V)\label{samplelagrangian}
\end{equation} where $\xi$ is the Fayet-Iliopoulos parameter.
The moduli space can be determined by taking the gauge coupling to be formally infinite,
i.e.~treating the linear superfield $V$ as an auxiliary field. Then $V$ is determined via
its equation of motion \begin{equation}
\sum_i q_i \Phi_i^* e^{2q_iV}\Phi_i + \xi =0, \label{sampledterm}
\end{equation} i.e.~$\Phi_i'=e^{q_iV}\Phi_i$ solve the usual D-term equation.
The \kahler potential of the moduli space is then
given by plugging the solution to \eqref{sampledterm} into \eqref{samplelagrangian}.
It can be generalized  to any gauge group, and the result agrees with the mathematical formula given in  Sec.~3.1 of \cite{BG}.
This shows that the \kahler potential is given just by the restriction of the original one
if $\xi=0$.
This analysis does not incorporate quantum corrections,
but it is well-known that for $\cN=2$ theories  the quantum effect does not modify the \hk structure, see Sec.~3 of \cite{Argyres:1996eh}.

\section{Geometry of the minimal nilpotent orbit}
\label{orbit}
Here we gather the relevant information on the \hk geometry of the minimal nilpotent orbit
of any simple group $G$, which coincides
with the centered moduli space of single instantons with gauge group $G$ \cite{Kronheimer,Br}.
We hope this section might be useful for anyone
who wants to deal with the one-instanton moduli space.
In the following $G$ stands for a compact simple Lie group,
$\frakg_\bR$ its Lie algebra.
We let $G_\bC$ and $\frakg_\bC$  be
complexifications of $G$ and $\frakg_\bR$ respectively.

The existence of a uniform description
of the one-instanton moduli space  applicable to any $G$
might be understood as follows:
we can construct a one-instanton configuration  easily by taking
a BPST instanton of $\SU(2)$ and regard it as an instanton of $G$
via a group embedding $\SU(2)\subset G$.
It is known that any one-instanton of $G$ arises in this manner \cite{Bernard:1977nr,Vainshtein:1981wh}.
The one-instanton moduli space is  then parameterized
by the position, the size, and the gauge orientation of the BPST instanton inside $G$.
This description realizes the one-instanton moduli space as a cone over a homogeneous manifold
$G/H$, where $H$ is the maximal subgroup of $G$ which commutes with the $\SU(2)$ used in the embedding.
It is however not directly suitable for the analysis of its complex structure.
For that purpose we use another realization of the one-instanton moduli space
as the minimal nilpotent orbit $\omin$ of $G$ \cite{Kronheimer}.

Let us define $\omin$.
First we decompose $\frakg_\bC$ into the Cartan generators $H^i$
and the raising/lowering operators $E_{\pm\rho}$ for roots $\rho$.
The minimal nilpotent orbit $\omin(G)$
is then the orbit of $(E_\theta)^*$ in $\frakg_\bC^*$,
where $\theta$ denotes the highest root: \begin{equation}
    \omin(G)=G_\bC \cdot (E_\theta)^* \subset \frakg_\bC^*.
\end{equation}
We will write $\omin$ without explicitly writing $G$ for the sake of simplicity when there is no confusion.

We think of elements of $\frakg_\bC$ as  holomorphic functions
on $\omin$, i.e.~we have holomorphic functions\footnote{More mathematically,
one has a natural holomorphic $\frakg_\bC^*$-valued
function $\bX:\omin\hookrightarrow \frakg_\bC^*$ given by the embedding.
Then every element $t\in \frakg_\bC$ gives a holomorphic function $(\bX,t)$ on $\omin$
via $x\in\omin \mapsto (\bX(x),t)$. Our $\bX^a$ is $(\bX,T^a)$ for a generator $T^a$ of $\frakg_\bC$.
We take a real basis of $\frakg_\bC$, so in fact
$T^a \in \frakg_\bR \subset \frakg_\bC$.
}\
$\bX^a$ ($a=1,\ldots,\dim G$) on $\omin$.
The defining equations of $\omin$ are a set of quadratic equations
which we call the Joseph relations \cite{Joseph}.\footnote{
Strictly speaking, the Joseph ideal is a two-sided ideal in the universal enveloping algebra of
$\frakg_\bC$, and what we use below is its associated ideal in the polynomial algebra. }\

These relations can be studied using a theorem of Kostant \cite{Garfinkle}:
Let $V(\alpha)$ denote the representation space of
a semisimple group $G$ with the highest weight $\alpha$,
and let $v\in V(\alpha)^*$ be a vector in the highest weight space.
The orbit $G_\bC\cdot v$ is then an affine algebraic variety whose defining ideal
$\cI$
is generated by its degree-two part $\cI_2$.
Furthermore, $\cI_2$ is given by the relation
\begin{equation} \label{i2rel}
\Sym^2 V(\alpha) = V(2\alpha)\oplus \cI_2
\end{equation} where we identify $\Sym^2 V(\alpha)$ as the space of
degree-two polynomials on $V(\alpha)^*$.
The minimal nilpotent orbit is exactly of this form
where $V(\alpha)$ is the adjoint representation, i.e.~
\begin{equation}
\omin=\{\bX\in \frakg_\bC^* \bigm| (\bX\otimes \bX)|_{\cI_2}=0 \}.
\end{equation}

For practice, let us apply this to the case $G=\SU(2)$. There, $V(\alpha)$ is the triplet representation,
so by \eqref{i2rel} $\cI_2$ is the singlet representation.
Therefore, if we parameterize $\mathfrak{su}(2)$ by $(a,b,c)$,
the minimal nilpotent orbit is given by the equation \begin{equation}
a^2+b^2+c^2=0,
\end{equation} which is $\bC^2/\bZ_2$ as it should be.

Now that we have given $\omin$ as a complex manifold,
let us describe its \hk structure.
The main fact we use is that $G$ acts isometrically on $\omin$, preserving the \hk structure.

There is a triplet of moment maps $\mu^a_s$
for this action where $a=1,\ldots, \dim G$ and $s=1,2,3$.
The functions $\bX^a$ are the holomorphic moment maps
of the $G$ action, i.e. $\bX^a=\mu^a_1 + i\mu^a_2$,
It follows that their Poisson bracket is \begin{equation}
[ \bX^a,\bX^b] = f^{ab}{}_c \bX^c
\end{equation} where $f^{ab}{}_c$ are the structure constants of $G$.
Phrased differently, the holomorphic symplectic structure underlying
the \hk structure of the nilpotent orbit
is the standard Kirilov--Kostant--Souriau symplectic form on the coadjoint orbit \cite{Kronheimer,Br}.

The conjugation is given by the $\SU(2)_R$
action, which sends $(\mu_1,\mu_2,\mu_3)$ to $(\mu_1,-\mu_2,-\mu_3)$.
Therefore \begin{equation}
\sigma^*(\bX^a) = (\bX^a)^*.
\end{equation}
The scaling dimension of $\bX$ is fixed to be two, as it should be
for the F-term in an $\cN=2$ supersymmetric theory.

Let us next describe a \kahler potential for $\omin$,
which was determined in \cite{KS1}.
The derivation boils down to the following:
$G$ acts on $\omin$ with cohomogeneity one, and by averaging over this action
we can consider $K$ to be $G$-invariant; so $K$ is a function
of $\tr \bX \bX^*$. $K$ should be of scaling dimension two,
so that $K$ is proportional to $\sqrt{\tr \bX \bX^*} $ up to a constant.
The constant factor can be fixed by considering a particular element on $\omin$.
For this purpose we again turn to the minimal nilpotent orbit of $\SU(2)$, which is
$\bC^2/\bZ_2$.
The normalization of the \kahler potential of the minimal nilpotent orbit of a general group
can then be determined because it contains the minimal nilpotent orbit of $\SU(2)$ as a subspace.

We parameterize $\bC^2$ by $(u,\tilde u)$ and
divide by the multiplication by $-1$.
We define our conventions for the holomorphic Poisson bracket and the \kahler potential
of a flat $\bH$ as follows:
\begin{equation}
K=|u|^2+|\tilde u|^2,\qquad
[u,\tilde u]=1.
\end{equation}
Now, $\bC^2/\bZ_2$ is parametrized by \begin{equation}
Z_{11}=u^2/2,\quad
Z_{12}=Z_{21}=u \tilde u /2 ,\quad
Z_{22}= \tilde u^2/2
\end{equation} which satisfy \begin{equation}
Z_{11} Z_{22}= Z_{12}{}^2.
\end{equation}
The \kahler potential is now \begin{equation}
K=2\sqrt{|Z_{11}|^2+|Z_{22}|^2+2|Z_{12}|^2}
= 2\sqrt{Z_{\alpha\beta} \bar Z ^{\alpha\beta}}.
\end{equation}
Then,  the moment map associated to the generator $J_3$ of non-R $\SU(2)$
acting on $\bC^2/\bZ_2$ can be explicitly calculated, with the result \begin{equation}
F=Z_{12}, \quad
D=\frac{2}{K}(|Z_{11}|^2-|Z_{22}|^2).
\end{equation}

Now that the preparation is done, we move on to the calculation of the Higgs branch
on both sides of the duality.

\section{$\SU(3)$ side}\label{SU3}

The theory has six quarks in the fundamental representation, \begin{equation}
Q^i_a,\qquad \tilde Q^a_i
\end{equation}where $a=1,\ldots,6$ and $i=1,2,3$.
As is well known, any $\SU(3)$-invariant polynomial constructed out of these fields
is a polynomial in the operators \cite{Argyres:1996eh}:
\begin{equation}
M^i{}_j= Q^i_a\tilde Q^a_j,\qquad
B^{ijk}=\epsilon^{abc} Q^i_a Q^j_b Q^k_c,\qquad
\tilde B_{ijk}=\epsilon_{abc} \tilde Q^a_i \tilde Q^b_j \tilde Q^c_k.
\end{equation}
In the following we study the Poisson brackets, the action of the conjugation, and the constraints
in turn.

\subsection{Poisson brackets}
The Poisson bracket of the basic fields is given by
\begin{equation}
[Q^i_a, \tilde Q_j^b] = \delta^i{}_j \delta^b{}_a.
\end{equation} Then we have, for example, \begin{equation}
[M^i{}_j, Q^k_a]=-\delta^k{}_j Q^i_a,
\end{equation} i.e.~$M^i{}_j$ is the generator of $\U(6)$.
We define $\tr M$ to be the trace of $M^i{}_j$, and
\begin{equation}
\hat M^i{}_j=M^i{}_j - \frac16\delta^i{}_j\tr M
\end{equation} is its traceless part.
$\hat M^{i}{}_j$ is the $\SU(6)$ generator and $\tr M$  the $\U(1)$ generator.
We define the $\U(1)$ charge $q$ of an operator $\cO$ to be given by \begin{equation}
[\tr M, \cO]=-q \cO.
\end{equation}
The most complicated bracket is
\begin{align}
[B^{ijk},\tilde B_{lmn}]&=18 M^{[i}{}_{[l} M^j{}_m \delta^{k]}{}_{n]} \nonumber \\
&= 18 \hat M^{[i}{}_{[l} \hat M^j{}_m \delta^{k]}{}_{n]}
+6 (\tr M) \hat M^{[i}{}_{[l} \delta ^j{}_m \delta^{k]}{}_{n]}
+\frac12 (\tr M)^2 \delta^{[i}{}_{[l} \delta ^j{}_m \delta^{k]}{}_{n]}. \label{BBpoisson}
\end{align}

\subsection{Conjugation}
We choose the involution on the elementary fields to be
\begin{equation}
\sigma^*(Q^i_a)=(\tilde Q^a_i)^*,\qquad
\sigma^*(\tilde Q^a_i)=-(Q^i_a)^*.
\end{equation}  Then the transformation of the  composites are \begin{align}
\sigma^*(M^i{}_j)&=-(M^j{}_i)^*, &
\sigma^*(\tr M)&= -(\tr M)^*,\\
\sigma^*(B^{ijk}) &= (\tilde B_{ijk})^*, &
\sigma^*(\tilde B_{ijk}) &= - (B^{ijk})^*.\label{Bconj}
\end{align}

\subsection{Constraints}

The constraints were studied in \cite{Argyres:1996eh}.
Those which come before imposing the F-term constraint are
\begin{align}
B^{ijk} \tilde B_{lmn} &= 6 M^{[i}{}_{l} M^j{}_m M^{k]}{}_n ,\label{BBMMM}\\
B^{ij[k}B^{lmn]} &=0, & \tilde B_{ij[k} \tilde B_{lmn]} &=0,\label{BB}\\
M^{[i}{}_j B^{klm]}&=0, &
M^{i}{}_{[j} \tilde B_{klm]}&=0.\label{MB}
\end{align}
The F-term constraint  \begin{equation}
Q^i_a \tilde Q^b_i - \frac13\delta^b_a (Q\tilde Q) =0
\end{equation} further imposes \begin{align}
\hat M^i{}_j B^{jkl}&=\frac16( \tr M) B^{ikl} , \label{MB3}\\
\hat M^i{}_j \tilde B_{ikl}&=\frac16 (\tr M)\tilde B_{jkl} ,\label{MB4} \\
\hat M^i{}_j M^j{}_k&=\frac16 (\tr M) M^i{}_k.\label{MM}
\end{align}
We will find it  convenient later to have constraints in terms of
irreducible representations (irreps) of $\SU(6)$.
We use the Dynkin labels to distinguish the irreps in the following.
The $MB=0$ relations \eqref{MB}, \eqref{MB3}, \eqref{MB4} give
\begin{align}
\hat M^{\{i}{}_l B^{[jk]\}l}&=0,&
\hat M^{\{i}{}_l \tilde B^{[jk]\}l}&=0, \label{MBi[jk]1}\\
\hat M^{l}{}_{\{i} B_{[jk]\}l}&=0,&
\hat M^{l}{}_{\{i} \tilde B_{[jk]\}l}&=0\label{MBi[jk]2}.
\end{align}
Here we defined the projector from a tensor with the structure $A_{i[jk]}$ to
the irrep $(1,1,0,0,0)$  by $A_{\{i[jk]\}} \equiv A_{i[jk]} -A_{[i[jk]]}$.
We also have
\begin{align}
\hat M^{[i}{}_l B^{jk]l}&=\frac16 (\tr M) B^{ijk}, &
\hat M^{l}{}_{[i} \tilde B_{jk]l}&=\frac16 (\tr M) \tilde B_{ijk}.\label{MB[ijk]}
\end{align}
The $MM=0$ relation \eqref{MM} gives \begin{align}
\hat M^{i}{}_j \hat M^j{}_k &= \frac16 \delta^i{}_k \hat M^{m}{}_n \hat M^n{}_m ,
\label{MMadjoint}\\
\hat M^{i}{}_j \hat M^j{}_i &= \frac16 (\tr M)^2.\label{MMsinglet}
\end{align}
The $BB=0$ relation \eqref{BB} gives \begin{equation}
B^{ikl}B_{jkl}=0,\qquad
\tilde B^{ikl}\tilde B_{jkl}=0.
\label{BBadj}
\end{equation}
Finally, the decomposition of the $B\tilde B=MMM$ relation gives,
using  \eqref{MMadjoint} and \eqref{MMsinglet}  repeatedly, \begin{align}
B^{ijk} \tilde B_{ijk} &= \frac29 (\tr M)^3,\label{BtBsinglet}\\
B^{ikl}\tilde B_{jkl} \proj{\adj} &= \frac29 (\tr M)^2 \hat M^i{}_j,\label{BtBadjoint}\\
B^{ijm}\tilde B_{klm} \proj{0,1,0,1,0} &=
    \frac23 (\tr M) \hat M^{[i}{}_{[k} \hat M^{j]}{}_{l]}  \proj{0,1,0,1,0},\label{BtB2}\\
B^{ijk}\tilde B_{lmn} \proj{0,0,2,0,0} & =
6 \hat M^{[i}{}_l \hat M^{j}{}_m \hat M^{k]}{}_n \proj{0,0,2,0,0}.\label{BtB3}
\end{align}

\section{Exceptional side}\label{E6}
\subsection{Poisson brackets}
We have chiral fields $\bX^a$ which transform in the adjoint of $E_6$,
and satisfy the quadratic Joseph identities.
We decompose $\bX^a$ under
the subgroup $\SU(2)\times \SU(6)\subset E_6$. It gives \begin{equation}
 X^i{}_j, \quad  Y^{[ijk]}_\alpha, \quad Z_{\alpha\beta}
\end{equation}where $X^i{}_j$ and $Z_{\alpha\beta}$ are the adjoints of $\SU(6)$ and
$\SU(2)$ respectively, and $Y^{ijk}_\alpha$ is in the doublet of $\SU(2)$ and in the representation $(0,0,1,0,0)$, i.e.~the three-index antisymmetric tensor, of $\SU(6)$.
The Poisson brackets of the fields $X$, $Y$ and $Z$ are exactly the Lie brackets
as explained above, which we take to be
\begin{align}
[X^i{}_j, X^k{}_l]&=\delta^i{}_l X^k{}_j -\delta^k{}_j X^i{}_l,\label{XXcommutator}\\
[Z_{\alpha\beta},Z_{\gamma\delta}] &= \frac12(
\epsilon_{\alpha\gamma} Z_{\beta\delta}+
\epsilon_{\beta\gamma} Z_{\alpha\delta}+
\epsilon_{\alpha\delta} Z_{\beta\gamma}+
\epsilon_{\beta\delta} Z_{\alpha\gamma})
\end{align}
and \begin{align}
[X^i{}_j, Y^{klm}_\alpha]&=-3  \delta^{[k}{}_{j} Y^{lm]i}_\alpha
+\frac12 \delta^i{}_j Y^{klm}_\alpha ,\\
[Z_{\alpha\beta}, Y^{ijk}_\gamma]&=\,  Y^{ijk}_{(\alpha}\epsilon_{\beta)\gamma}
\end{align} and finally \begin{equation}
[Y^{ijk}_\alpha,Y^{lmn}_\beta]=
\epsilon^{ijklmn} Z_{\alpha\beta}
-\frac32 \,  \epsilon_{\alpha\beta}
(X^{[i}{}_p \epsilon^{jk]lmnp}+X^{[l}{}_p \epsilon^{mn]ijkp}).
\end{equation} The final commutation relation can also be written as \begin{equation}
[Y^{ijk}_\alpha, Y_{lmn\,\beta}]=-18 X^{[i}{}_{[l} \delta ^j{}_m \delta^{k]}{}_{n]}
-6 Z_{\alpha\beta}\delta^{[i}{}_{[l} \delta ^j{}_m \delta^{k]}{}_{n]}.
\end{equation}

As we explained above, $X$, $Y$ and $Z$ are the holomorphic moment maps
of the $E_6$ action.
Therefore the contribution from $\omin$ to the F-term constraint for the $\SU(2)$ gauge group
is given just by $Z_{\alpha\beta}$.

We take the bracket of $v$ and $\tilde v$ to be \begin{equation}
[v_\alpha,\tilde v_\beta]=\epsilon_{\alpha\beta}.
\end{equation}
Then we have \begin{equation}
[v_{(\alpha}\tilde v_{\beta)},v_\gamma]=v_{(\alpha}\epsilon_{\beta)\gamma},
\end{equation} and
\begin{equation}
[(v\tilde v),v_\alpha]=v_\alpha,\qquad
[(v\tilde v),\tilde v_\alpha]=-\tilde v_\alpha.
\end{equation}
Recall that we define $
(uw)\equiv u_\alpha w_\beta\epsilon^{\alpha\beta}
$ for two doublets $u_\alpha$ and $w_\alpha$.
It is straightforward to check
that $v_{(\alpha}\tilde v_{\beta)}$ is the moment map of the $\SU(2)$ action on $v$ and $\tilde v$.
Thus the F-term condition is  \begin{equation}
v_{(\alpha}\tilde v_{\beta)}+Z_{\alpha\beta}=0.\label{E6Fterm}
\end{equation}

\subsection{Conjugation}
We take the conjugation on the variables $v,\tilde v$ to be \begin{equation}
\sigma^*(v_\alpha)=(\tilde v_\beta)^* \epsilon_{\alpha\beta},\qquad
\sigma^*(\tilde v_\alpha)=(v_\beta)^* \epsilon_{\alpha\beta}.
\end{equation}

In terms of our variables $(X^i{}_j,Y^{ijk}_\alpha,Z_{\alpha\beta})$,
the conjugation acts as follows:
 \begin{align}
\sigma^*(X^i{}_j)&=-(X^j{}_i)^*,\\
\sigma^*(Y^{ijk}_\alpha)&=(Y_{ijk\, \beta})^* \epsilon_{\alpha\beta}, &
\sigma^*(Y_{ijk\, \alpha})&=-(Y^{ijk}_\beta)^* \epsilon_{\alpha\beta}, \\
\sigma^*(Z_{\alpha\beta})&=(Z_{\gamma\delta})^* \epsilon_{\alpha\gamma}
\epsilon_{\beta\delta}.
\end{align}

\subsection{Constraints}

As explained in Sec.~\ref{orbit}, the Joseph relations are given by \begin{equation}
(\bX \otimes \bX)|_{\cI_2}=0
\end{equation} where $\cI_2$ is given by the relation \begin{equation}
\Sym^2 V(\adj) = V(2\adj)\oplus \cI_2.
\end{equation}
Here, $V(\adj)$  is the adjoint representation of $E_6$
whose Dynkin label is 
$\adj=\Esix010000$.
We then have \begin{equation}
\cI_2 =  V\left(\Esix100001\right)\oplus V\left(\Esix000000\right).
\end{equation}

\begin{table}[b]
\[
\begin{array}{c@{\ }|@{\ }c@{\ }|@{\ \,}c@{\ \,}|c|cccccccc}
& \SU(2)& \SU(6) & \,\text{in $\cI_2$}\, & \,\text{in $\Sym^2 V(\adj)$}\\
\hline
1. & \mathbf{3} &(1,0,0,0,1) & 1 & 2 \\
2. & \mathbf{2}&(1,1,0,0,0)& 1& 1\\
3. & \mathbf{2}&(0,0,0,1,1) & 1&1 \\
4. & \mathbf{2}& (0,0,1,0,0) & 1 & 2\\
5. & \mathbf{1}&(0,1,0,1,0) & 1 & 2\\
6. & \mathbf{1}&(1,0,0,0,1) & 1&1 \\
7. & \mathbf{1}& (0,0,0,0,0) &2& 3
\end{array}
\]
\caption{Decomposition of $\cI_2$ in terms of $\SU(2)\times\SU(6) \subset E_6$.
\label{decomposition}}
\end{table}

The representations which appear in $\cI_2$,
decomposed under $\SU(2)\times \SU(6)$,
are summarized in Table~\ref{decomposition}.
The table reads as follows: e.g.~for relation 4,
the fourth column tells us
there is one Joseph identity transforming as a doublet in $\SU(2)$ and as
$(0,0,1,0,0)$ under $\SU(6)$,
but the fifth column says one can construct two objects in this representation
from bilinears in $X^i{}_j$, $Y^{ijk}_\alpha$ and $Z_{\alpha\beta}$.
This means the identity has the form\begin{equation}
0=Y^{ijk}_\alpha Z_{\beta\gamma}\epsilon^{\alpha\beta} + c_4 X^{[i}{}_l Y^{jk]l}_\gamma,
\end{equation} where $c_4$ needs to be fixed,
which can be done e.g.~by explicitly evaluating
the right hand  side on a few elements on the nilpotent orbit.
Elements on the nilpotent orbit can be readily generated,
because one knows that the point \begin{equation}
X^i{}_j=0,\qquad Y^{ijk}_\alpha=0,\qquad Z_{11}=1, \qquad Z_{12}=Z_{22}=0
\end{equation} is on the nilpotent orbit by definition. Then the rest of the points
can be generated by the coadjoint action of $E_6$, which can be obtained by exponentiating the structure constants.

Carrying out this program, we obtain the following full set of Joseph identities:
\begin{align}
1. \qquad 0&= X^i{}_j  Z_{\alpha\beta} + \frac14 Y^{ikl}_{(\alpha} Y_{jkl \beta)},\label{J1}\\
2. \qquad 0& =X^l{}_{\{i} Y_{[jk]\}l\alpha},\label{J2}\\
3. \qquad 0& =X^{\{i}{}_l Y^{[jk]\}l}_\alpha,\label{J3}\\
4. \qquad 0& = Y^{ijk}_\alpha Z_{\beta\gamma}\epsilon^{\alpha\beta} +  X^{[i}{}_l Y^{jk]l}_\gamma,
\label{J4}\\
5. \qquad 0& = (Y^{ijm}_\alpha Y_{klm\beta}\epsilon^{\alpha\beta} -4  X^{[i}{}_{[k} X^{j]}{}_{l]})\proj{0,1,0,1,0}  ,\label{J5} \\
6. \qquad 0 & = X^i{}_k X^{k}{}_j -\frac16\delta^i{}_j X^k{}_l X^l{}_k,\label{J6}\\
7. \qquad 0 &= Y^{ijk}_\alpha Y_{ijk\beta} \epsilon^{\alpha\beta}
+ 24 Z_{\alpha\beta} Z_{\gamma\delta} \epsilon^{\alpha\gamma}\epsilon^{\beta\delta} ,\label{J7}\\
\text{7'.} \qquad 0 &= X^i{}_j X^j{}_i +   3 Z_{\alpha\beta} Z_{\gamma\delta} \epsilon^{\alpha\gamma}\epsilon^{\beta\delta}.\label{J7p}
\end{align}

\subsection{Gauge invariant operators}\label{5.4}
Let us enumerate the generators of the $\SU(2)$-invariant
operators constructed out of
$v_\alpha$, $\tilde v_\alpha$, and $X^i{}_j$, $Y^{ijk}_\alpha$, $Z_{\alpha\beta}$,
using the F-term equation \eqref{E6Fterm} and the Joseph identities \eqref{J1} $\sim$ \eqref{J7p}.
Suppose we have a monomial constructed from those fields.
We first replace every appearance of $Z_{\alpha\beta}$ by $-v_{(\alpha}\tilde v_{\beta)}$.
All the $\SU(2)$ indices are contracted by  epsilon tensors of $\SU(2)$.
Therefore the monomial is a product of $X^i{}_j$, $(v\tilde v)$, $(Y^{ijk}v)$, $(Y^{ijk}\tilde v)$
and $(Y^{ijk} Y^{lmn})$. The last of these can be eliminated using the Joseph identities. Indeed,
the combination of the relations \eqref{J5}, \eqref{J7} and \eqref{J7p} gives
a Joseph identity of the form \begin{equation}
Y^{ijk}_\alpha Y_{lmn\,\beta}\epsilon^{\alpha\beta}= 18 X^{[i}{}_{[l} X^j{}_m \delta^{k]}{}_{n]}
-3 Z_{\alpha\beta} Z_{\gamma\delta}\epsilon^{\alpha\gamma}\epsilon^{\beta\delta}
\delta^{[i}{}_{[l} \delta ^j{}_m \delta^{k]}{}_{n]}. \label{JJ}
\end{equation}
We conclude that any $\SU(2)$-invariant polynomial is a polynomial in \begin{equation}
X^i{}_j,\quad  (v\tilde v),\quad (Y^{ijk}v),\quad\text{and}\quad (Y^{ijk}\tilde v).
\end{equation}

\section{Comparison}\label{comparison}
\subsection{Identification of operators}
Let us now proceed to the comparison of
the structures we studied in Sec.~\ref{SU3} and in Sec.~\ref{E6}.
We first make the following identification: \begin{equation}
\hat M^i{}_j = X^i{}_j,\qquad \tr M = -3(v\tilde v).
\end{equation}  These are the moment maps of the flavor symmetries $\SU(6)$ and $\U(1)$,
so the identification is fixed including the coefficients, and then
the Poisson brackets involving either $\hat M$ or $\tr M$ automatically agree.
The conjugation acting on $X^i{}_j$, $(v\tilde v)$ also agrees with that on
$\hat M^i{}_j$ and $\tr M$.

We then set
\begin{equation}
B^{ijk} = c (Y^{ijk} v),\qquad
\tilde B_{ijk} = \tilde c (Y_{ijk}\tilde v). \label{Bident}
\end{equation}
One has \begin{equation}
\sigma((Y^{ijk} v))
=(Y_{ijk}\tilde v)^*.
\end{equation} To be consistent with \eqref{Bconj},
we need to have
\begin{equation}
\tilde c=c^*. \label{cc}
\end{equation}
Let us then calculate the Poisson bracket of $(Y^{ijk}v)$ and $(Y_{lmn}\tilde v)$
using \eqref{JJ}.
We have \begin{equation}
[(Y^{ijk}v),(Y_{lmn}\tilde v)]=-18 X^{[i}{}_{[l} X^j{}_m \delta^{k]}{}_{n]}
 +18  (v \tilde v) X^{[i}{}_{[l} X^j{}_m \delta^{k]}{}_{n]}
 -\frac{9}{2}(v \tilde v)^2.
\end{equation} Comparing with the bracket $[B^{ijk},\tilde B_{lmn}]$ calculated
in \eqref{BBpoisson}, we find they indeed agree if
$c\tilde c=-1$.
Thus we conclude $c=\tilde c=i$, i.e.~ \begin{equation}
B^{ijk}=i (Y^{ijk}v),\qquad
\tilde B_{ijk}=i (Y_{ijk} \tilde v).
\end{equation}

\subsection{Constraints}

Now, let us check using the Joseph relations that the constraints on the $\SU(3)$ side, listed in Eqs.~\eqref{MBi[jk]1} $\sim$ \eqref{BtB3}, can be correctly reproduced  on the exceptional side.
\begin{itemize}
\item \eqref{MBi[jk]1}: Contract $v$ or $\tilde v$ to the relation 2, \eqref{J2}.
\item \eqref{MBi[jk]2}: Contract $v$ or $\tilde v$ to the relation 3, \eqref{J3}.
\item \eqref{MB[ijk]}:  Contract $v$ or $\tilde v$ to the relation 4, \eqref{J4}.
\item \eqref{MMadjoint}: This is exactly the relation 6, \eqref{J6}.
\item \eqref{MMsinglet}: This is exactly the relation 7', \eqref{J7p}.
\item \eqref{BBadj}:  Contract  $v_\alpha v_\beta$
or $\tilde v_\alpha \tilde v_\beta$  to the relation 1 \eqref{J1}.
\end{itemize}
As for the relation of the type $B\tilde B=MMM$,
\begin{itemize}
\item \eqref{BtBsinglet}: The singlet part. Contract  $v_\alpha\tilde v_\beta$ to  the relation 7 \eqref{J7}.
\item \eqref{BtBadjoint}: The adjoint part. Contract  $v_\alpha\tilde v_\beta$ to  the relation 1 \eqref{J1}.
\item \eqref{BtB2}:  The $(0,1,0,1,0)$ part. Contract  $v_\alpha\tilde v_\beta$ to the relation 5 \eqref{J5}.
\item \eqref{BtB3}:
This is the $(0,0,2,0,0)$ part and is slightly trickier,  but it follows from a cubic Joseph identity
\begin{equation}
0=\epsilon^{\alpha\gamma}\epsilon^{\beta\delta}Z_{\alpha\beta}Y^{ijk}_{\gamma}Y_{lmn,\delta}\proj{0,0,2,0,0}-6 X^{[i}{}_l X^{j}{}_m X^{k]}{}_n\proj{0,0,2,0,0}
\end{equation}  upon replacing $Z_{\alpha\beta}$ with $v_{(\alpha}\tilde v_{\beta)}$.
This cubic  Joseph identity itself can be derived from the
quadratic Joseph identities, as it should be.
First, we use the relation 4 \eqref{J4} to show \begin{equation}
\epsilon^{\alpha\gamma}\epsilon^{\beta\delta} Z_{\alpha\beta} Y^{ijk}_{\gamma} Y_{lmn\,\delta}
\proj{0,0,2,0,0}
\propto X^{[i}{}_pY^{jk]p}_\alpha Y_{lmn\,\beta}\epsilon^{\alpha\beta}.
\end{equation}
 Now, the antisymmetric product of two $Y$'s contain both
the singlet and the $(0,1,0,1,0)$ part. One sees the singlet drops out inside the projector
to the $(0,0,2,0,0)$ part, so we have \begin{equation}
\propto \left(X^{[i}{}_p (Y^{jk]p} Y_{lmn})\proj{0,1,0,1,0}\right)\proj{0,0,2,0,0}.
\end{equation} Then we use the relation 5 \eqref{J5}  to transform this to \begin{equation}
\propto X^i{}_l X^{j}{}_m X^{k}{}_n\proj{0,0,2,0,0}.
\end{equation} The proportionality constant can be fixed, e.g.~by evaluating
on a few points on the orbit. This concludes the comparison of the constraints.
\end{itemize}

\section{Conclusions}\label{conclusion}
In the previous three sections, we determined the Higgs branches both on the
$\SU(3)$ side and on the exceptional side. We demonstrated that their defining equations
agree, and furthermore exhibited that the Poisson bracket and the conjugation are
the same on both sides. As was stated in Sec.~\ref{formalities} and will be detailed in Appendix~\ref{twistors}, these are (almost) sufficient
to conclude that they are the same as \hk manifolds. To remove any remaining doubts,
we compare the \kahler potentials of the two sides in Appendix~\ref{kahler}. Again,
they show remarkable agreement with one another.

Thus we definitely showed the agreement of the Higgs branches
of the new $S$-duality pair proposed  by Argyres and Seiberg in \cite{Argyres:2007cn},
which provides a convincing check of their conjecture.
In this paper we only dealt with the example involving $E_6$, but
there are more examples of similar dualities in \cite{Argyres:2007cn} and \cite{Argyres:2007tq}.
It would be interesting to carry out the same analysis of the Higgs branches
to those examples.
A pressing issue is to understand the Argyres-Seiberg duality more fully.
For example, it would be nicer
to have an embedding of this duality in string/M-theory.
We hope to revisit these problems in the future.

\section*{Acknowledgments}
The authors thank Alfred D. Shapere for collaboration at an early
stage of the project. They would also like to thank S. Cherkis, C.
R. LeBrun, H. Nakajima for helpful discussions. They also relied
heavily on the softwares LiE\footnote{ It can be downloaded from
\texttt{http://www-math.univ-poitiers.fr/\~{}maavl/LiE/}.} and
Mathematica.  DG is supported in part by the DOE grant
DE-FG02-90ER40542 and in part by the Roger Dashen membership in the
Institute for Advanced Study. AN was supported in part by the Martin A.\ and Helen
Chooljian Membership at the Institute for Advanced Study, and in part by the
NSF under grant numbers PHY-0503584 and PHY-0804450.
YT is supported in part by the NSF grant PHY-0503584,
and in part by the Marvin L. Goldberger membership in the Institute
for Advanced Study.

\appendix

\section{Conventions}\label{conventions}

Greek indices $\alpha,\beta$ are for the doublets of $\SU(2)$,
$a,b,c,\ldots$ for the triplets of $\SU(3)$ and
$i,j,k,\ldots$ for the sextets of $\SU(6)$.
We define \begin{equation}
(uw)\equiv u_\alpha w_\beta\epsilon^{\alpha\beta}
\end{equation}for two doublets $u_\alpha$ and $w_\alpha$,

We use the following sign conventions for the epsilon tensors
of $\SU(2)$, $\SU(3)$ and $\SU(6)$:
\begin{equation}
\epsilon^{\alpha\beta}=-\epsilon_{\alpha\beta},
\quad \epsilon^{abc} = \epsilon_{abc},
\quad \epsilon^{ijklmn} = \epsilon_{ijklmn}.
\end{equation}
We normalize the antisymmetrizer $[abc...]$ and
the symmetrizer $(abc...)$ so that they are projectors, i.e.~\begin{equation}
T^{ijk}=T^{[ijk]}
\end{equation} for the antisymmetric tensors $T^{ijk}$, etc.
We raise and lower three antisymmetrized indices of $\SU(6)$ via the following rule:
\begin{equation}
T^{ijk}= \frac16\epsilon^{ijklmn} T_{lmn}, \qquad
T_{lmn}= \frac16 T^{ijk}\epsilon_{ijklmn}.
\end{equation}

Our convention for the placement of the indices of the
complex conjugate is e.g. \begin{equation}
\bar Z^{\alpha\beta} \equiv (Z_{\alpha\beta})^*,
\end{equation}  i.e.~the complex conjugation is always accompanied by
the exchange of subscripts and superscripts, as is suitable for the action of $\SU$ groups.

We take the \kahler potential of a flat $\bC$ parameterized by $z$
with the standard metric to be \begin{equation}
K=|z|^2.
\end{equation}

\section{Twistor spaces of \hk cones}\label{twistors}

Recall that a \hk manifold $\cM$ admits a continuous family of complex structures $J_\zeta$,
parameterized by $\zeta \in \bC\bP^1$.
The full information in the \hk metric is captured by this family of complex structures
and their Poisson brackets.
It can be encoded into purely holomorphic data on a complex manifold
$\cZ$, the \textit{twistor space} of $\cM$, as we now review.

Topologically $\cZ = \cM \times \bC\bP^1$.  Its complex structure can be specified by
specifying which functions on $\cZ$ are holomorphic:  they are
$f(x,\zeta)$ which are holomorphic in $\zeta$ for fixed $x\in\cM$,
and also holomorphic in $x$ with
respect to complex structure $J_\zeta$ for fixed $\zeta$.  Hence we may view $\cZ$ as a holomorphic
fiber bundle over $\bC\bP^1$, where the fiber over $\zeta$ is just a copy of $\cM$, equipped with
complex structure $J_\zeta$.

The Poisson brackets on the holomorphic functions in each fiber glue
together globally to give a bracket operation on $\cZ$.  This bracket operation
is globally twisted by the line bundle $\cO(-2)$:  i.e.~given local holomorphic functions $f_1$, $f_2$ we
get a local section
$\{f_1,f_2\}$ of $\cO(-2)$, and more generally if $f_1$, $f_2$ are sections of $\cO(d_1)$, $\cO(d_2)$
then $\{f_1,f_2\}$ is a section of $\cO(d_1 + d_2 - 2)$.
Finally there is
an involution $\sigma$ on $\cZ$, simply defined by $(x, \zeta) \to (x, - 1 / \bar\zeta)$.  This is
an antiholomorphic involution, since the complex structure $J_\zeta$ is opposite to $J_{-1/\bar\zeta}$.

As a complex manifold $\cZ$ is a fibration over $\bC\bP^1$, and
$(x,\zeta)$ with $x$ fixed gives a holomorphic section of this fibration,
which is invariant under $\sigma$.
The normal bundle to this section is isomorphic to the line bundle $\cO(1)^{\oplus n}$
where $n$ is the complex dimension of $\cM$.
Conversely, a holomorphic section of $\cZ$ which is invariant under $\sigma$
and whose normal bundle is isomorphic to $\cO(1)^{\oplus n}$
is called a twistor line.
Therefore, the points on $\cM$ give rise to a $n$-dimensional family of twistor lines on $\cZ$.

It was shown in \cite{Hitchin:1986ea} that given $\cZ$, together with its
Poisson brackets and antiholomorphic involution, one can
canonically reconstruct a \hk metric on the space of twistor lines.
Therefore, to check  that our two \hk cones are the same is essentially
to check that their twistor spaces $\cZ$ are the same.

Now, the twistor space of a \hk cone can be constructed
from the data we  described in Sec.~\ref{formalities}, i.e.~the
Poisson bracket, the dilatation and the conjugation on $\cM$.
We pick one complex structure induced from the \hk structure,
and regard $\cM$ as a complex manifold.
We then form $\cZ$ as a complex manifold as \begin{equation}
\cZ= ( ( \bC^2\setminus (0,0) ) \times \cM ) / \bC^\times
\end{equation} where $\bC^\times$ acts on the first factor by multiplication,
and on the second factor as the natural complexification
of the action of the dilatation.
Then the Poisson bracket  on $\cM$ naturally induces one on $\cZ$.
We define $\sigma$ on $\cZ$ to send $(z,w,x)\in \bC^2 \times \cM$
to $(-\bar w,\bar z, \sigma(x))$.
Then it is straightforward to check that this $\cZ$ is the twistor space
of $\cM$, using the $\SU(2)_R$ action on $\cM$ rotating three complex structures.

There is a subtle problem remaining, however.
Namely, the theorem in \cite{Hitchin:1986ea}  asserts that there is a component of the space of the twistor lines of $\cZ$ which agrees metrically with the original \hk manifold $\cM$,
but does not exclude the possibility that the space of twistor lines
has many components, each of which is a \hk manifold with the same complex structure but
with a different metric. Mathematicians the authors consulted know no concrete example where
this latter possibility is realized, so the authors think it quite unlikely that  our two \hk manifolds
are the same as holomorphic symplectic manifolds but not as \hk manifolds.
To dispel this last possibility, in the next Appendix we directly compare the \kahler potential
of our two \hk manifolds.

\section{Comparison of the \kahler potential}\label{kahler}
In this Appendix, we describe the method to calculate and compare the \kahler potential
of the Higgs branches on the two sides of the duality.

\subsection{Exceptional side}
The invariant norm of $E_6$ in our notation is \begin{equation}
Z_{\alpha\beta} \bar Z^{\alpha\beta}
+\frac16 Y^{ijk}_\alpha \bar Y^\alpha_{ijk}
+X^i{}_j \bar X^j{}_i.
\end{equation}
Therefore the correctly normalized \kahler potential is \begin{equation}
K_{E_6}=2\sqrt{
Z_{\alpha\beta} \bar Z^{\alpha\beta}
+\frac16 Y^{ijk}_\alpha \bar Y^\alpha_{ijk}
+X^i{}_j \bar X^j{}_i },
\end{equation}  and the D-term for the $\SU(2)\subset E_6$ is \begin{equation}
D^{(E_6)}_{\alpha\beta}=\frac{2}{K_{E6}}
\left[
    Z_{\alpha\gamma} \bar Z^{\gamma\delta} \epsilon_{\delta \beta}
+   Z_{\beta\gamma} \bar Z^{\gamma\delta} \epsilon_{\delta \alpha}
+\frac1{12}(
    Y^{ijk}_\alpha \bar Y_{ijk}^\gamma \epsilon_{\gamma\beta}
+   Y^{ijk}_\beta \bar Y_{ijk}^\gamma \epsilon_{\gamma\alpha}
)
\right].
\end{equation}

We also have quarks $v_\alpha$, $\tilde v_\alpha$
which have \begin{equation}
K_{v,\tilde v} = \sum_\alpha (|v_\alpha|^2 + |\tilde v_\alpha|^2)
\end{equation} and \begin{equation}
D^{(v,\tilde v)}_{\alpha\beta} = \frac12(
v_\alpha \bar v^\gamma \epsilon_{\gamma\beta}
+v_\beta \bar v^\gamma \epsilon_{\gamma\alpha}
+\tilde v_\alpha \bar{\tilde  v}^\gamma \epsilon_{\gamma\beta}
+\tilde v_\beta \bar{\tilde  v}^\gamma \epsilon_{\gamma\alpha}
).
\end{equation}

The \kahler potential of the exceptional side is thus given by \begin{equation}
K_{v,\tilde v} + K_{E_6} \label{E6sideKahler}
\end{equation}  restricted to the locus \begin{equation}
v_{(\alpha} \tilde v_{\beta)} + Z_{\alpha\beta}=0,\qquad
D^{(v,\tilde v)}_{\alpha\beta}+D^{(E_6)}_{\alpha\beta}=0
\end{equation}
expressed as a function of $M^i_j$, $B^{ijk}$, $\tilde B^{ijk}$
and their complex conjugates.

\subsection{$\SU(3)$ side}
We start from the \kahler potential \begin{equation}
K=\sum_{i,a} | Q^i_a| ^2 + | \tilde Q^a_i| ^2.
\end{equation}
Using the analysis in \cite{Argyres:1996eh}, the \kahler potential on the quotient was determined in
\cite{Antoniadis:1996ra} as
\begin{equation}
K=2\sum_{i=1,2,3} \sqrt{m_i^2+\frac{\nu^2}4} \label{SU(3)sideKahler}
\end{equation}
where $(m_1^2, m_2^2,m_3^2, 0,0,0)$ are the eigenvalues of
$M^i{}_j \bar M^j{}_k$, and $\nu$ is defined by\begin{equation}
3\nu=\sum_{i,a} | Q^i_a| ^2 - | \tilde Q^a_i| ^2,
\end{equation} i.e.~1/3 of the $\U(1)$ D-term. In terms of gauge invariants
we have \begin{align}
\prod_{i=1,2,3} \left(
\sqrt{m_i^2+\frac{\nu^2}4} + \frac\nu 2
\right) &= \frac16 B^{ijk }\bar B_{ijk}, \label{nuB} \\
\prod_{i=1,2,3} \left(
\sqrt{m_i^2+\frac{\nu^2}4} - \frac\nu 2
\right) &= \frac16 \tilde B^{ijk }\bar {\tilde B}_{ijk}.\label{nuBb}
\end{align}

\subsection{Comparison}
Now, the \kahler potentials of the two sides, \eqref{E6sideKahler}
and \eqref{SU(3)sideKahler} should agree as functions of
$M$, $B$ and $\tilde B$, but we have not been able to check that analytically.
Instead, one can check it numerically on as many points on the quotient
as computer time allows.
The algorithm is as follows: \begin{enumerate}
\item Generate a point $\bX=(X^i{}_j,Y^{ijk}_\alpha,Z_{\alpha\beta})$
 on the nilpotent orbit of $E_6$, by applying an element of the group $E_6$
 to the point $(Z_{11},Z_{12},Z_{22})=(1,0,0)$, $X^i{}_j=Y^{ijk}_\alpha=0$.
\item Find $v_\alpha,\tilde v_\alpha$ which satisfy \begin{equation}
v_{(\alpha} \tilde v_{\beta)} + Z_{\alpha\beta}=0.
\end{equation} This is more or less unique up to $\bC^\times$ action
on $v,\tilde v$.
\item Apply $\SL(2,\bC)$ action to $(v,\tilde v, \bX)$
to find the solution of the D-term equation, \begin{equation}
D^{(v,\tilde v)}_{\alpha\beta}+D^{(E_6)}_{\alpha\beta}=0.
\end{equation}  This is equivalent to the minimization of \begin{equation}
K_{v,\tilde v}(g(v),g(\tilde v))+ K_{E6} (g(\bX))
\end{equation} where $g$ is an $\SL(2,\bC)$ action.
\item Form $M$, $B$, $\tilde B$ from $v$, $\tilde v$ and $\bX$
thus obtained, and calculate $\nu$ and $m_i$.
At this point, two checks of the sanity of the calculation are possible.
One is to see that three eigenvalues of $M\bar M$  are zero.
Another is to see that $\nu$ determined from \eqref{nuB}, \eqref{nuBb} is equal to
\begin{equation}
\nu = \sum_{\alpha} |v_\alpha|^2 - |\tilde v_\alpha|^2.
\end{equation} The latter fact follows from the identification of $\nu$ as
1/3 of the $\U(1)$ moment map on the quotient.
\item Evaluate the \kahler potential of the $\SU(3)$ side using \eqref{SU(3)sideKahler}
and compare it to that of the exceptional side \eqref{E6sideKahler}.
\end{enumerate}
We implemented the algorithm above in Mathematica,
and found that the value of the \kahler potential
at any points agrees on both sides of the duality to arbitrary accuracy.\footnote{We thank H. Elvang
for improvement of the accuracy in the calculation.}\
An analytic proof of the agreement of the \kahler potential will be welcomed.

\section{Mathematical Summary}\label{math-summary}
Let us summarize briefly in the language of mathematics what was done in this paper.
Let $M(m,n)$ be \begin{equation}
M(m,n)=\mathrm{Hom}(V,W) \oplus \mathrm{Hom}(W,V)
\qquad \text{where}\  V=\bC^m,\quad W=\bC^n
\end{equation} which is a flat \hk space of quaternionic dimension $mn$.
It has a natural triholomorphic action of $\U(m)\times \U(n)$ induced from its action on $V$ and $W$.
Let $N(m,n)$ be the flat \hk space \begin{equation}
N(m,n)=\bR^m \otimes_\bR \bH^n
\end{equation} of quaternionic dimension $mn$,
which has a natural triholomorphic action of $\SO(m)\times\Sp(n)$.

One then defines a \hk quotient $A$ by \begin{equation}
A_1=M(6,3)\hkq \SU(3).
\end{equation}
We consider another \hk quotient \begin{equation}
A_2=(N(2,1) \times \omin(E_6) )\hkq \Sp(1)
\end{equation} where $\omin(G)$ is the minimal nilpotent orbit of the group $G$,
and the $\Sp(1)$ action on $\omin(E_6)$ is given by considering the
maximally compact subgroup $\Sp(1)\times \SU(6) \subset E_6$.

One sees easily that $A_{1,2}$ are both of quaternionic dimension $10$,
 both carry a natural triholomorphic action of $\SU(6)\times \U(1)$.
Our claim is that $A_1=A_2$ as \hk cones. We demonstrated that $A_1$ and $A_2$ match
as holomorphic symplectic varieties by explicitly showing that their defining equations and
the holomorphic symplectic forms are the same. We also found that the twistor spaces of
$A_1$ and $A_2$ are the same as complex manifolds with antiholomorphic involution,
but could not show that $A_1$ and $A_2$ correspond to the same family of twistor lines.
Instead we directly compared the \kahler potentials of $A_1$ and $A_2$. Again
we could not rigorously prove the equivalence, but we performed numerical
calculations of the \kahler potential which convinced us that they agree.

The equivalence of $A_{1,2}$ was suggested  by the analysis of a
new type of $S$-duality in four-dimensional $\cN=2$ supersymmetric gauge theories in \cite{Argyres:2007cn}.  In \cite{Argyres:2007cn,Argyres:2007tq}, more examples of the same type of duality were described,
of which we record two more here.

Now consider \begin{equation}
B_1= N(12,2) \hkq \Sp(2)
\end{equation} and \begin{equation}
B_2 = \omin(E_7) \hkq \Sp(1).
\end{equation} Here $\Sp(1)$ acts on $\omin(E_7)$ through the maximal subgroup
$\Sp(1)\times \SO(12)\subset E_7$.
The quaternionic dimension of $B_{1,2}$ is 14, and
both have triholomorphic actions of $\SO(12)$.
We believe $B_1=B_2$ as \hk cones.

For an example which involves $\omin(E_8)$, consider \begin{equation}
C_1= ( Z \oplus N(11,3) ) \hkq \Sp(3).
\end{equation} Here $Z$ is a pseudoreal irreducible representation of $\Sp(3)$
of quaternionic dimension 7, which arises as \begin{equation}
\wedge^3_\bC \, X = Z\oplus X
\end{equation} where $X=\bC^6$  is the defining representation of $\Sp(3)$.
Let us  take another \hk quotient \begin{equation}
C_2 = \omin(E_8)\hkq \SO(5)
\end{equation} where $\SO(5)$ acts via embedding \begin{equation}
\SO(5)\times \SO(11) \subset \SO(16)\subset E_8.
\end{equation} It is easy to check that $C_{1,2}$ are both of quaternionic dimension 19,
and $\SO(11)$ acts triholomorphically on both $C_1$ and $C_2$.
We predict  that $C_1=C_2$ as \hk cones.

\def\url#1{\href{#1}{#1}}
\bibliographystyle{utphys}
\bibliography{bib}{}
\end{document}